\documentclass[11pt,preprint]{aastex}

\shorttitle{Angular diameters of giant stars from the CHARA Array}
\shortauthors{Baines et al.}

\usepackage{amsmath}
\begin{document}


\title{Angular Diameters and Effective Temperatures of Twenty-five K Giant Stars from the CHARA Array}

\author{Ellyn K. Baines$^\dagger$}
\affil{Remote Sensing Division, Naval Research Laboratory, 4555 Overlook Avenue SW, \\ Washington, DC 20375}
\email{ellyn.baines.ctr@nrl.navy.mil}
\altaffiltext{$^\dagger$}{Some of the observations described here was completed while with the Center for High Angular Resolution Astronomy, Georgia State University, P.O. Box 3969, Atlanta, GA 30302-3969.}

\author{Michaela P. D\"ollinger}
\affil{ESO, Karl-Schwarzschild-Strasse 2, D-85748 Garching bei M\"unchen, Germany}

\author{Felice Cusano, Eike W. Guenther, Artie P. Hatzes}
\affil{Th\"uringer Landessternwarte Tautenburg, Sternwarte 5, D-07778 Tautenburg, Germany}

\author{Harold A. McAlister, Theo A. ten Brummelaar, Nils~H.~Turner, Judit Sturmann, \\ Laszlo Sturmann, P. J. Goldfinger, Christopher D. Farrington}
\affil{Center for High Angular Resolution Astronomy, Georgia State University, P.O. Box 3969, \\ Atlanta, GA 30302-3969}

\author{Stephen T. Ridgway}
\affil{Kitt Peak National Observatory, National Optical Astronomy Observatory, P.O. Box 26732, Tucson, AZ 85726-6732} 

\begin{abstract}

Using Georgia State University's CHARA Array interferometer, we measured angular diameters for 25 giant stars, six of which host exoplanets. The combination of these measurements and \emph{Hipparcos} parallaxes produce physical linear radii for the sample. Except for two outliers, our values match angular diameters and physical radii estimated using photometric methods to within the associated errors with the advantage that our uncertainties are significantly lower. We also calculated the effective temperatures for the stars using the newly-measured diameters. Our values do not match those derived from spectroscopic observations as well, perhaps due to the inherent properties of the methods used or because of a missing source of extinction in the stellar models that would affect the spectroscopic temperatures.

\end{abstract}


\section{Introduction}

Giant star radii have been measured in the past using various interferometers, including the Mark III \citep[85 giants and supergiants,][]{2003AJ....126.2502M}, the Palomar Testbed Interferometer \citep[69 giants and supergiants,][]{1999AJ....117..521V}, the Navy Prototype Optical Interferometer \citep[50 giants and supergiants,][]{1999AJ....118.3032N}, and the Center for High Angular Resolution Astronomy (CHARA) Array \citep[4 Hyades giants,][]{2009ApJ...691.1243B}. These measurements are valuable because these are the stars populating the coolest, most luminous part of the Hertzsprung-Russell (H-R) diagram \citep{1999AJ....117..521V}. What makes the sample of giant stars under consideration here particularly interesting is that they are potential exoplanet hosts, and planetary candidates have been discovered around six of the stars already. 

Two important characteristics of a star are its mass and radius. For giant stars, the determination
of these parameters is indirect and heavily model
dependent. In practice, spectroscopic observations to 
measure the surface gravities (log~$g$), effective temperatures ($T_{\rm eff}$), and iron abundances ([Fe/H]) can be combined with a distance measurement to derive the stellar radius. Fitting evolutionary tracks to the position of the star in the H-R diagram then yields the mass.
The reliability of these measurements depend both on the validity of the
model atmospheres and the stellar evolution code. Unfortunately this is an
uncertain process because the evolutionary tracks of stars with a wide range of
masses all converge to near the same region of the H-R diagram as they
evolve up the giant branch. In particular the mass estimates derived from
evolutionary tracks depend critically on several parameters
hidden in the tracks, such as the mixing length parameter and its assumed
constancy for all stars, the unknown helium content in the core, and uncertainties about the nature of the convection zone. As a result, using
different tracks can produce different masses, and in the absence of good
calibrating objects no set of tracks can be claimed to provide the
best results. On the other hand, if one can test and calibrate these 
evolutionary tracks by comparing the theoretically-determined mass and radius
to observed values, then one can have some faith in applying these
tracks to stars for which direct measurements of these stellar parameters
is not possible.

A star's mass is not only important for its evolution, but it
should play an important role in the type of planetary system a star will
form. There are a number of Doppler surveys searching for planets 
around evolved giant stars with stellar masses of 1 to 2 $M_\odot$
\citep[e.g.,][]{2009ApJ...693..276N,doellinger07,2005A&A...437L..31S,2005PASJ...57...97S}. All are plagued by the same problem in that they rely on evolutionary tracks to determine the stellar mass. Until these are calibrated both the mass of the host star and the planet are uncertain.

A more reliable means of calculating the stellar
mass independent of evolutionary tracks and model atmospheres is
using stellar oscillation observations, as the frequency of stellar oscillations is related
to the mean density of the star. If one has an accurate stellar radius it is
simple to compute a stellar mass from the oscillation
frequencies that is model independent. Depending on the accuracy of the diameter measurements, the masses can be
measured to an accuracy of $\sim$2$\%$ \citep{2009A&A...494..237T} to $\sim$15$\%$ 
\citep{hat07}.

There is increasing evidence that most and possibly all giant stars
show stellar oscillations \citep[e.g.,][]{2006A&A...448..689D,fra02,hat94}, which are due to p-mode oscillations where pressure is the restoring force. Thus giant stars are an ideal class of objects for deriving fundamental stellar parameters. They are abundant, they have large angular diameters suitable for interferometric measurements, and they exhibit stellar oscillations with radial velocity amplitudes of a few to several tens of m/s, which are easily measurable by state-of-the-art techniques. The observed oscillation frequencies constrain the internal structure of the star \citep{2006ApJ...647..558B} and interferometry measures the star's size, and the combination leads to the mass of the star. Once stellar isochrones have been refined and calibrated for these evolved stars, they can be used to determine the masses of all planet-hosting giant stars. Because collecting data
on the oscillation frequencies requires considerable telescope resources and can only be done for
relatively few stars, we first present our results on interferometric measurements on a larger sample of giant stars.

The advantage interferometry provides is the ability to directly measure stellar angular diameters. Once the angular diameters are known for these giant stars, physical radii and effective temperatures can be calculated when combined with other parameters, such as the parallax, bolometric flux, interstellar absorption, and bolometric corrections. The radii and effective temperatures are important values that characterize the parent star as well as the environment in which the exoplanet resides for those stars hosting planets. Section 2 describes the spectroscopic measurements of $T_{\rm eff}$ and log~$g$ for the sample, $\S$3 discusses the interferometric observations, $\S$4 explains how the angular diameters, linear radii, and $T_{\rm eff}$ were determined, and $\S$5 explores the physical implications of the interferometric observations.

\section{Spectroscopic observations}

Our sample of K giant stars were obtained from the planet search survey
of \citet{doellinger07}. As part of this program the $T_{\rm eff}$ and log~$g$ were 
measured, which allowed us to estimate the stellar radii and masses. Table~\ref{spec} lists the 25 stars observed here, and planets have already been found orbiting HD~73108 \citep{doellinger07}, HD~139357 and HD~170693 \citep{doellinger09a}, HD~32518 and HD~136726 \citep{doellinger09b}, and HD~167042 \citep{2008ApJ...675..784J,2008PASJ...60.1317S,doellinger09c}. Three additional stars show long-period variations in their radial velocity measurements: HD~106574, HD~157681, and HD~200205 \citep{doellinger09d}. The targets chosen for our observing list are bright ($V < 6.5$) giant stars that showed significant short-term variability indicative of stellar pulsations, which made them excellent candidates for both stellar oscillation observations and interferometric measurements.

The spectroscopic observations were carried out using the Coud\'e
\'Echelle spectrograph of the 2-m-Alfred Jensch telescope of the
Th\"uringer Landessternwarte Tautenburg. The spectrograph has a
resolving power of $\Delta\lambda/\lambda =67000$ and the wavelength
range used was 4700 to 7400 \AA.  Standard IRAF routines were used for
subtracting the bias offset, flat-fielding, subtracting the scattered
light, extracting the spectra, and for the wavelength calibration\footnote{IRAF is distributed by the National Optical Astronomy Observatories, which are operated by the Association of Universities for Research in Astronomy, Inc., under cooperative agreement with the National Science Foundation.}.

In order to determine the stellar parameters from the spectra, a grid
of model atmospheres from \citet{gustafsson75} was used in which a
plane-parallel atmosphere in local thermodynamic equilibrium was
assumed. We selected 144 unblended Fe\,I and 8 Fe\,II lines in the
wavelength range 5806 and 6858 \AA\, using the line list of
\citet{pasquini04}.  The iron abundance [Fe/H] was determined by
assuming that Fe\,I lines of different equivalent widths have to give
the same relative abundance of iron.  For the effective temperature, an excitation equilibrium of Fe\,I and Fe\,II for lines of
different excitation potentials was used, and the surface gravity was determined from the ionization balance of Fe\,I to Fe\,II
lines \citep{doellinger08}.  The resulting [Fe/H], $T_{\rm eff}$, and log~$g$ values are listed in Table~\ref{spec}.

\section{Interferometric observations}
Interferometric observations were obtained using the CHARA Array, a six element optical-infrared interferometer located on Mount Wilson, California \citep{ten05}. All
observations used the pupil-plane ``CHARA Classic''
beam combiner in the $K'$-band at 2.15~$\mu$m while visible
wavelengths (470-800 nm) were used for tracking and tip/tilt
corrections. The observing procedure and data reduction process
employed here are described in \citet{mca05}.

We interleaved calibrator and target star observations so that every target was flanked by calibrator observations made as close in time as possible, which allowed us to convert instrumental target and calibrator visibilities to calibrated
visibilities for the target. Reliable calibrators were chosen to be single
stars with expected visibility amplitudes $>$85$\%$ so they were nearly
unresolved on the baselines used, which meant uncertainties in the
calibrator's diameter did not affect the target's diameter calculation
as much as if the calibrator star had a significant angular size. In a few cases, a calibrator had a stellar companion but at such a distance that light from the secondary star would not contaminate our interferometric measurements and the calibrator could therefore be treated as a single star.

To check for possible unseen close companions that would contaminate our observations, we created spectral energy
distribution (SED) fits based on published $UBVRIJHK$ photometric
values obtained from the literature for each calibrator to establish
diameter estimates. This also allowed us to see if there was any
excess emission associated with a low-mass stellar companion or
circumstellar disk. Calibrator candidates showing displaying
variable radial velocities or any other indication of companions were
discarded.

We used Kurucz model atmospheres\footnote{Available to download at
  http://kurucz.cfa.harvard.edu.}  based on $T_{\rm eff}$ and log~$g$
values to calculate limb-darkened angular diameters for the
calibrators. The stellar models were fit to observed photometry after
converting magnitudes to fluxes using \citet{col96} for $UBVRI$ values
and \citet{coh03} for $JHK$ values. See Table \ref{observations} for
the $T_{\rm eff}$ and log~$g$ used and the resulting limb-darkened
angular diameters.

\section{Determination of angular diameter and $T_{\rm eff}$}
\begin{figure}[h]
\includegraphics[width=1.0\textwidth, angle=90]{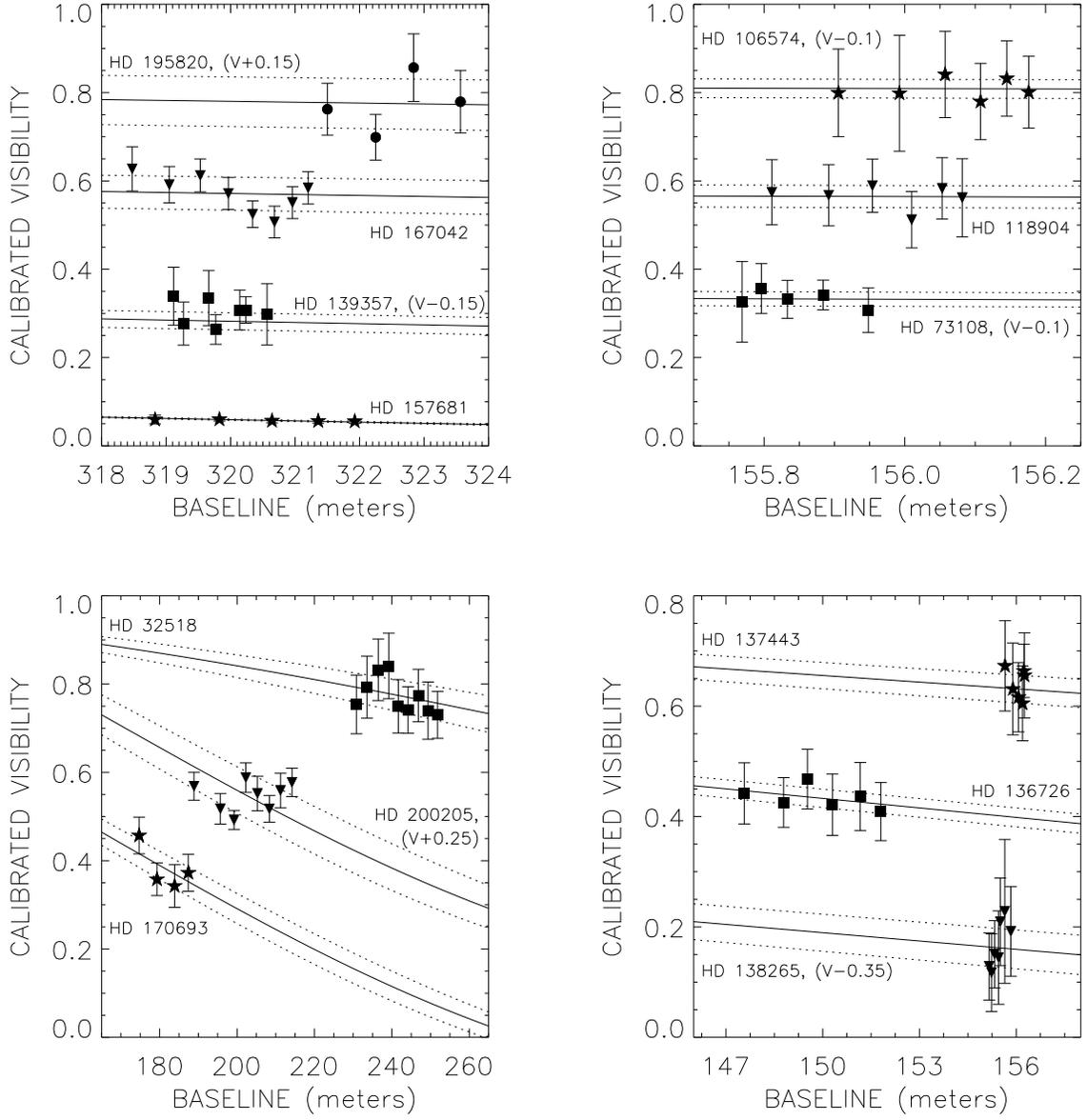}
\caption{LD disk diameter fits for all the stars observed with one calibrator except HD 214868. The solid line represents the theoretical visibility curve for a star with the best fit $\theta_{\rm LD}$, the dashed lines are the 1$\sigma$ error limits of the diameter fit, the solid symbols are the calibrated visibilities, and the vertical lines are the measured errors. Some of the stars' visibilities were shifted as indicated by ``(V $\pm$ $\#$)'' so they would not overlap other data points.}
  \label{stack_all}
\end{figure}

\begin{figure}[h]
\includegraphics[width=1.0\textwidth]{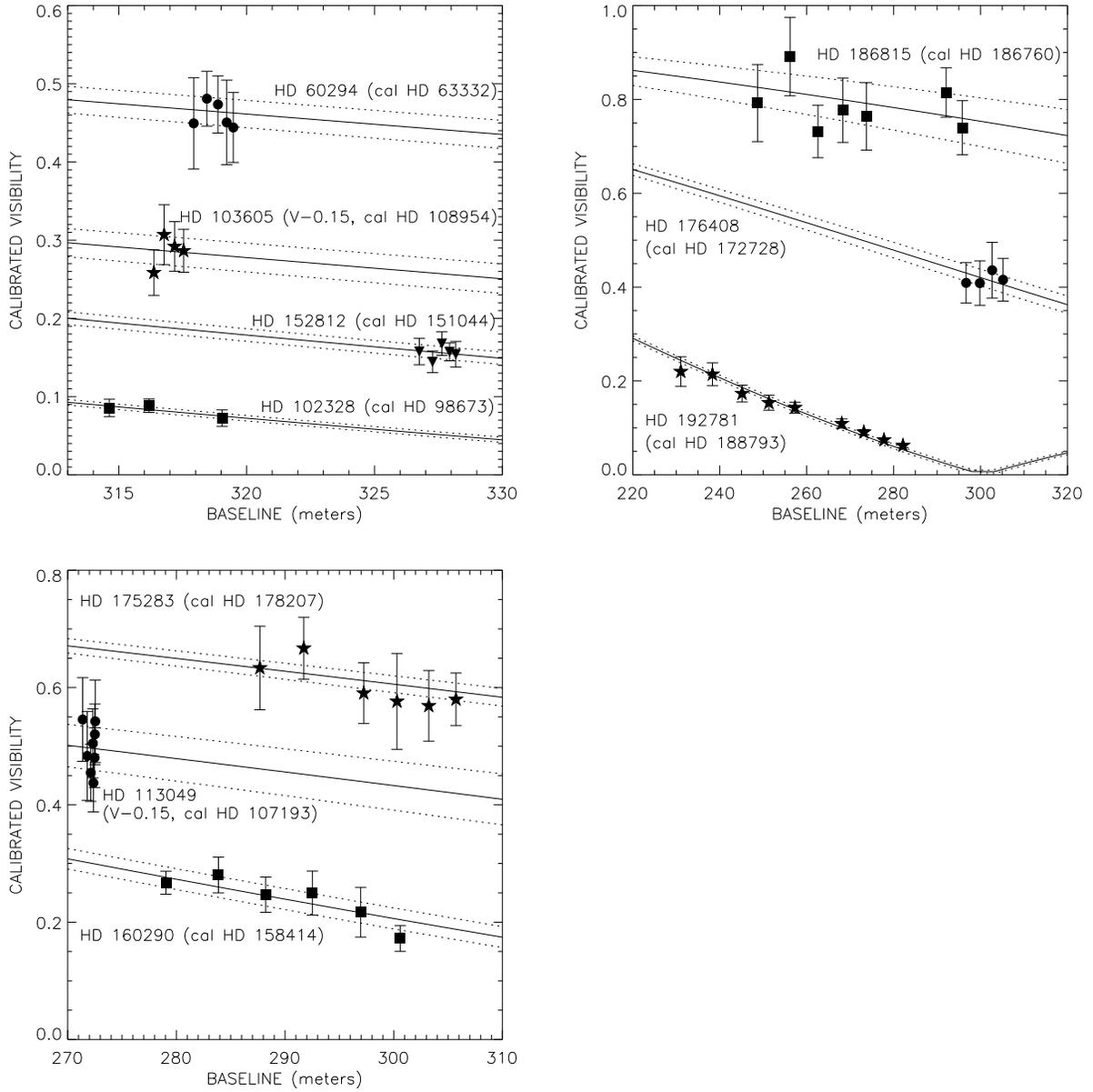}
\caption{LD disk diameter fits for all the stars observed with two calibrators except HD 150010. The symbols are the same as listed in Figure~1. For the sake of clarity, the data points for one calibrator only are shown.}
  \label{stack_all2}
\end{figure}

\begin{figure}[h]
\includegraphics[width=0.6\textwidth]{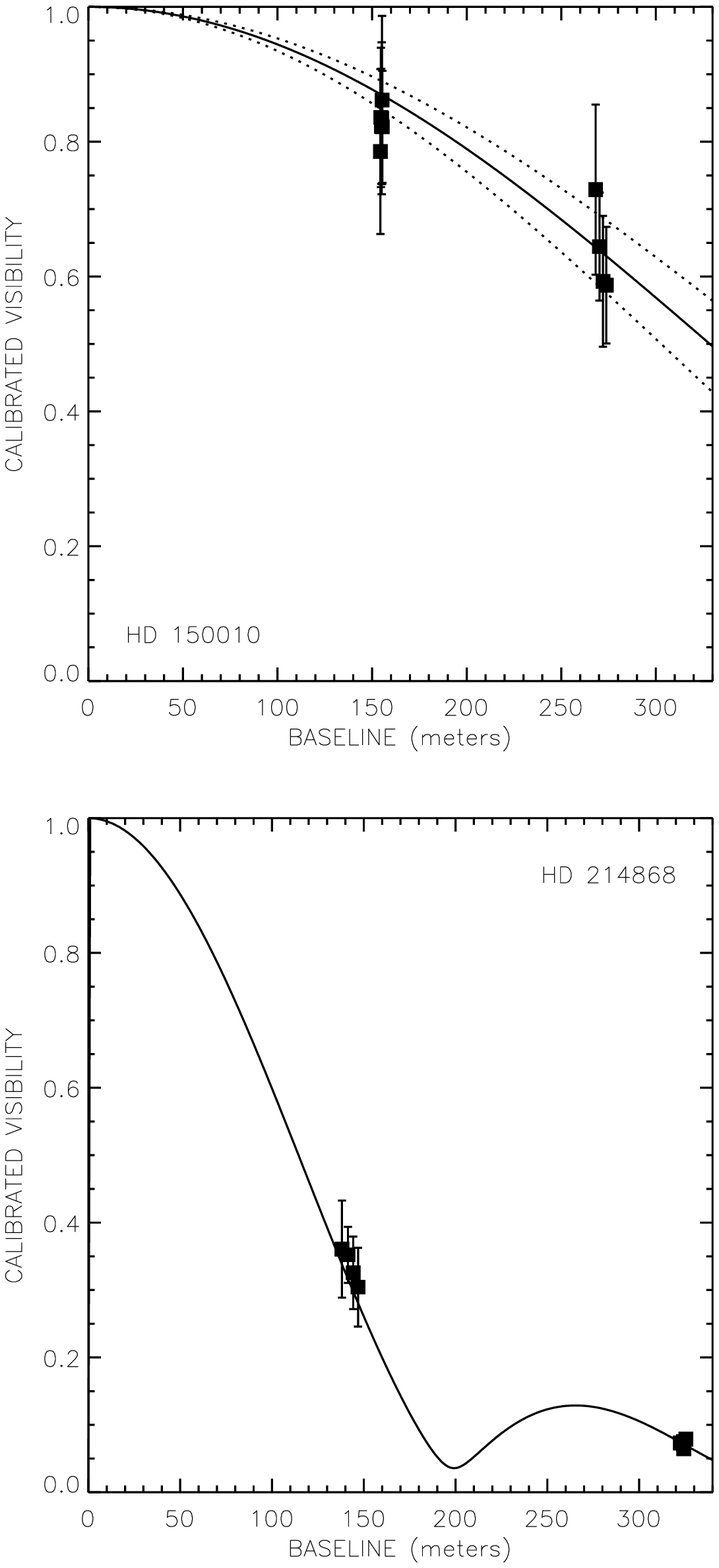}
\caption{LD disk diameter fits for HD 150010 (top panel) and HD 214868 (bottom panel). The symbols are the same as listed in Figure~\ref{stack_all}.}
  \label{others}
\end{figure}

The observed quantity of an interferometer is defined as the visibility ($V$), which is fit to a model of a uniformly-illuminated disk (UD) that represents the observed face of the star. Diameter fits to $V$ were based upon the UD approximation given by $V = [2 J_1(x)] / x$, where $J_1$ is the first-order Bessel function and $x = \pi B \theta_{\rm UD} \lambda^{-1}$, where $B$ is the projected baseline at the star's position, $\theta_{\rm UD}$ is the apparent UD angular diameter of the star, and $\lambda$ is the effective wavelength of the observation \citep{shao92}. A more realistic model of a star's disk involves limb-darkening (LD), and relationship incorporating the linear limb darkening coefficient $\mu_{\lambda}$ \citep{han74} is:
\begin{equation}
V = \left( {1-\mu_\lambda \over 2} + {\mu_\lambda \over 3} \right)^{-1}
\times
\left[(1-\mu_\lambda) {J_1(\rm x) \over \rm x} + \mu_\lambda {\left( \frac{\pi}{2} \right)^{1/2} \frac{J_{3/2}(\rm x)}{\rm x^{3/2}}} \right] .
\end{equation}
Table~\ref{calib_visy} lists the Modified Julian Date
(MJD), projected baseline ($B$) at the time of observation, projected baseline position angle
($\Theta$), calibrated visibility ($V_c$), and error in $V_c$ ($\sigma
V_{c}$) for each giant star observed.  Figures \ref{stack_all} through \ref{others} show the LD diameter fits for all the stars.

The limb-darkening coefficient was obtained from \citet{cla95} after
adopting the $T_{\rm eff}$ and log~$g$ values required for each star
observed.  The resulting LD angular diameters are listed in
Table~\ref{inf}.  The average difference between the UD and
LD diameters are on the order of a few percent, and the final angular
diameters are little affected by the choice of $\mu_{\lambda}$. All but four stars have $\theta_{\rm LD}$ errors of 2$\%$ or less, three of the four have errors of only 3$\%$, and the final star has a 5$\%$ error. Additionally,
the combination of the interferometric measurement of the star's
angular diameter plus the \emph{Hipparcos} parallax \citep{2007hnrr.book.....V} allowed us
to determine the star's physical radius. The results are also listed in Table~\ref{inf}. In principle, one can calculate the mass of each star from the physical radius and log $g$ values. However, the formal errors in log $g$ lead to errors in such mass estimates near the 50$\%$ level, thereby significantly decreasing their usefulness to this analysis.

For each $\theta_{\rm LD}$ fit, the errors were derived via the
reduced $\chi^2$ minimization method \citep{2003psa..book.....W,1992nrca.book.....P}: the diameter fit with the lowest
$\chi^2$ was found and the corresponding diameter was the final
$\theta_{\rm LD}$ for the star. The errors were calculated by finding
the diameter at $\chi^2 + 1$ on either side of the minimum $\chi^2$
and determining the difference between the $\chi^2$ diameter and
$\chi^2 +1$ diameter. In calculating the diameter errors in Table~\ref{inf}, 
we adjusted the estimated visibility errors to force the reduced $\chi^2$ to 
unity because when this is omitted, the reduced $\chi^2$ is well under 1.0, 
indicating we are overestimating the errors in our calibrated visibilities.

Limb-darkened angular diameters were estimated using the relationship described in \citet{2004AandA...426..297K} between the ($V-K$) color and log $\theta_{\rm LD}$ (see $\theta_{\rm estimate}$ in Table~\ref{spec}). The table also lists $R_{\rm estimate}$, which were derived using $\theta_{\rm estimate}$ and the stars' parallaxes. The major weakness of this method lies in the uncertainties surrounding the $K$-magnitudes, which were taken from two sources: \emph{The Two-Micron Sky Survey} \citep[TMSS, ][errors $\sim$2-5$\%$]{1969tmss.book.....N} and \emph{The 2MASS All-Sky Catalog of Point Sources} \citep[2MASS, ][errors $\sim$6-12$\%$]{2003tmc..book.....C}. Preference was given to the former because 2MASS measurements saturate at magnitudes brighter than $\sim$3.5 in the $K$-band even when using the shortest exposure time\footnote{Explanatory Supplement to the 2MASS All Sky Data Release and Extended Mission Products, http://www.ipac.caltech.edu/2mass/releases/allsky/doc/.}. The large errors associated with 2MASS magnitudes for these bright stars led to large errors in angular diameter and physical radii estimates.

Once $\theta_{\rm LD}$ was determined interferometrically, the $T_{\rm eff}$ was calculated using the relation 
\begin{equation}
F_{\rm BOL} = {1 \over 4} \theta_{\rm LD}^2 \sigma T_{\rm eff}^4,
\end{equation}
where $F_{\rm BOL}$ is the bolometric flux and $\sigma$ is the Stefan-Bolzmann constant. The stars' $V$ and $K$ magnitudes were dereddened using the extinction curve described in \citet{1989ApJ...345..245C} and interstellar absorption ($A_{\rm V}$) values were from \citet{2005A&A...430..165F} except for HD~113049 and HD~176408, which had no $A_{\rm V}$ in the literature. $A_{\rm V}$ values for these two stars were estimated through a non-linear, least squares fit and a reddening prescription from \citet{1999PASP..111...63F}, who presented a wavelength-dependent extinction curve. The intrinsic broad-band color ($V-K$) was calculated and bolometric corrections (BCs) were determined by interpolating between the [Fe/H] = +0.2, 0.0, and -1.0 tables found in \citet{1999A&AS..140..261A}. They point out that in the range of 6000 K $\geq T_{\rm eff} \geq$ 4000 K, their BC calibration is symmetrically distributed around a $\pm$0.10 magnitude band when compared to other calibrations. The average BC used here is 0.55, and because 0.10 is 18$\%$ of 0.55, we assigned a 18$\%$ error bar to our BC values. The bolometric flux was determined by applying the BC for each star and the $T_{\rm eff}$ was calculated (see Table~\ref{inf}). All $T_{\rm eff}$ errors are $\leq 4 \%$, 11 stars have errors of $\leq 2 \%$, and the major source of error in calculating $T_{\rm eff}$ stemmed, again, from uncertainties in $K$-magnitudes.

Giant star masses were estimated using the PARAM stellar model\footnote{http://stev.oapd.inaf.it/cgi-bin/param$\_$1.0} from \citet{2000A&AS..141..371G} with a modified version of the method described in \citet{2006A&A...458..609D}. The input parameters for each star were its interferometrically-measured $T_{\rm eff}$, its spectroscopically-derived [Fe/H], its $V$ magnitude from \citet{Mermilliod}, and its Hipparcos parallax \citep{2007hnrr.book.....V} along with the corresponding error for each value. The model used these inputs to estimate each star's age, mass, radius, $(B-V)_0$, and log~$g$ using the isochrones and a Bayesian estimating method, calculating the probability density function separately for each property in question. da Silva et al. qualify mass estimates as ``more uncertain'' than other properties, so the resulting masses listed in Table~\ref{spec} should be viewed as a rough estimates only.

\section{Results and discussion}

\begin{figure}[h]
\includegraphics[width=1.0\textwidth]{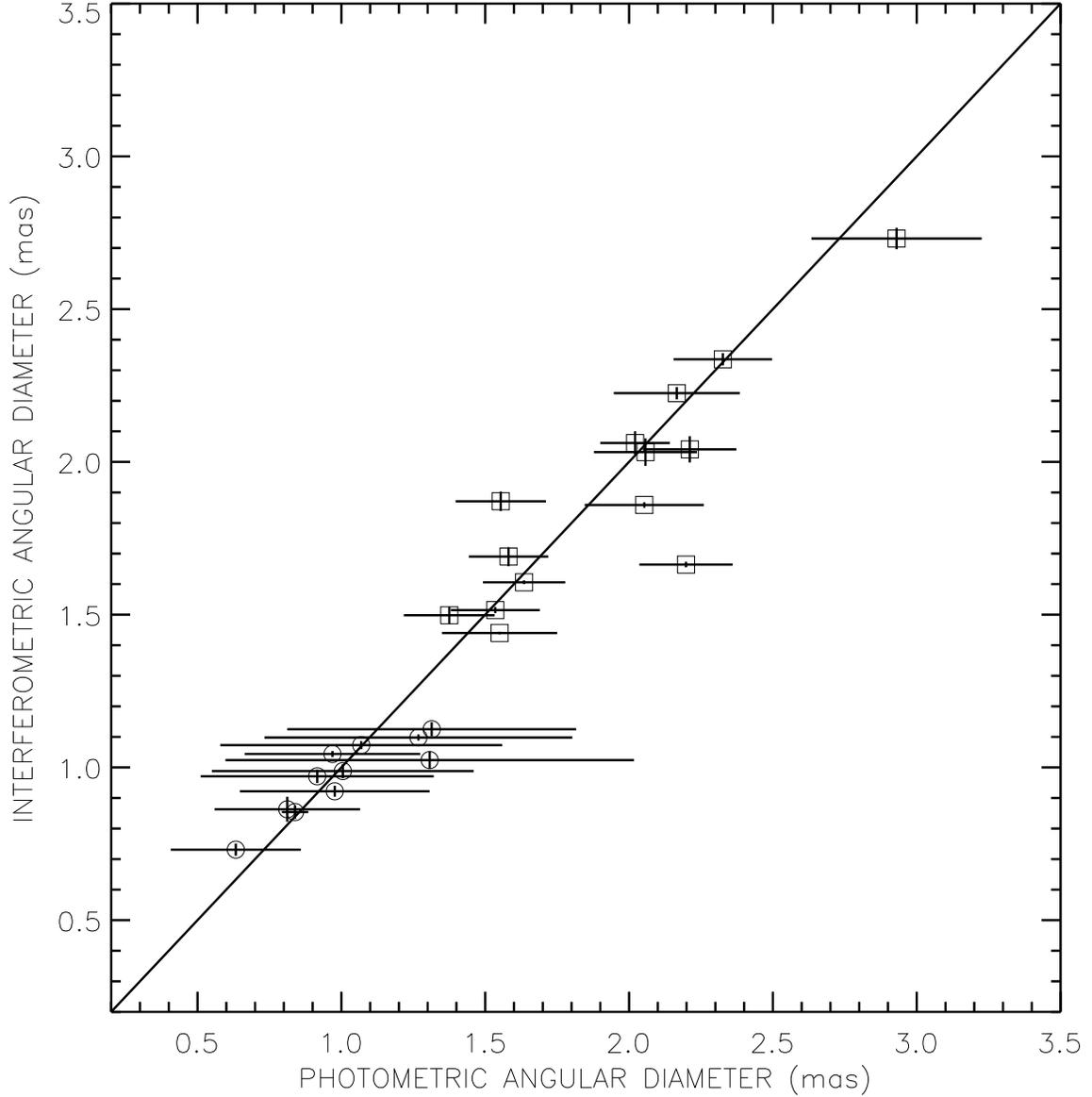}
\caption{A comparison of photometrically-estimated and interferometrically-measured diameters. The squares and circles represent diameters estimated using $K$ magnitudes from TMSS and 2MASS, respectively, and the diagonal solid line indicates a 1:1 ratio for the diameters. Note the significantly larger error bars associated with the photometric diameters, particularly those using 2MASS data. The outliers above and below the line are HD~118904 and HD~157681, respectively, and the discrepancies may be due to the calibrator used (see $\S$5 for more details).}
  \label{calcobs}
\end{figure}

\begin{figure}[h]
\includegraphics[width=1.0\textwidth]{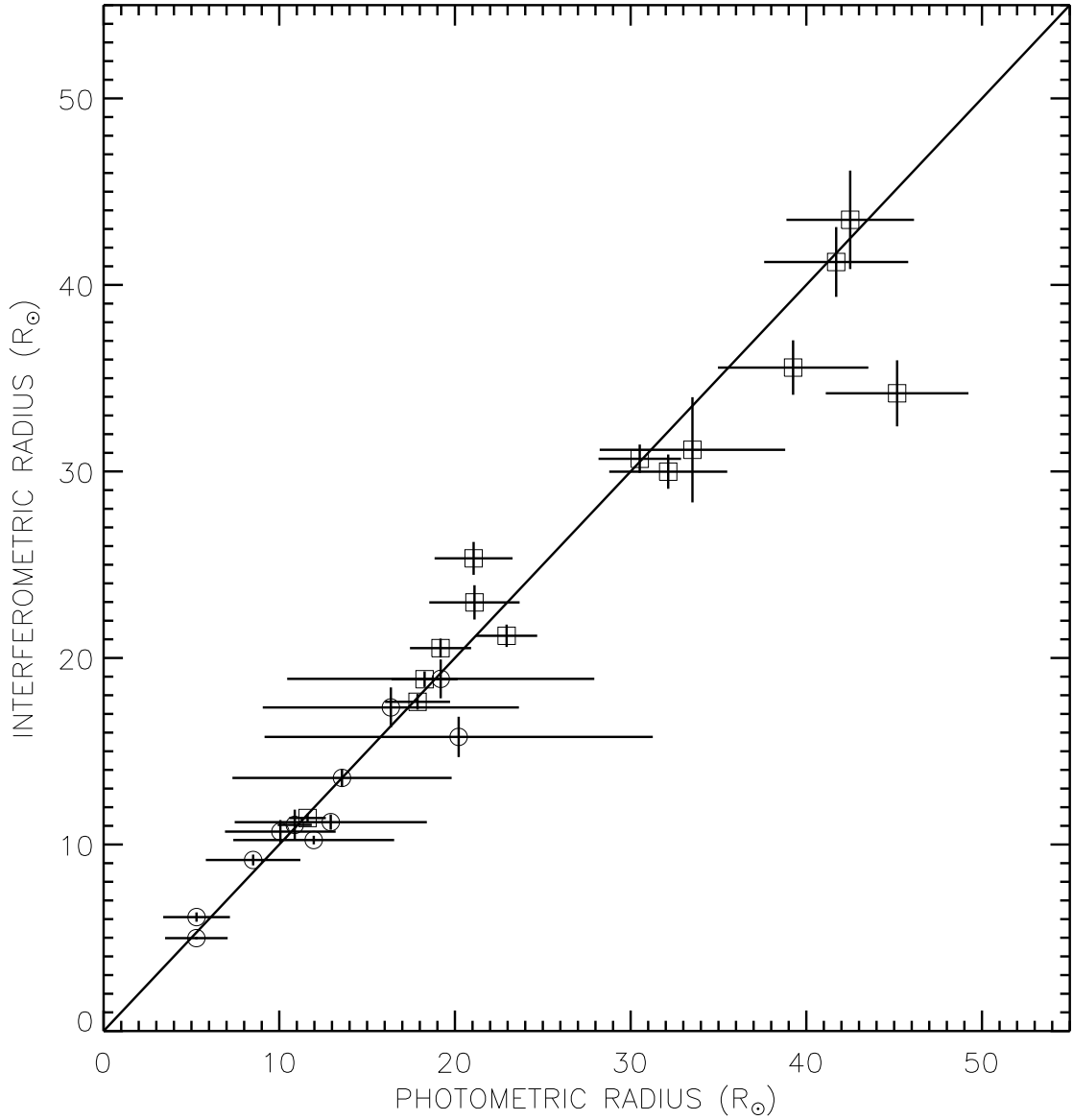}
\caption{A comparison of photometrically- and interferometrically-determined linear radii. The symbols and outliers are the same as listed in Figure~\ref{calcobs}.}
  \label{radii}
\end{figure}

In order to check how well the estimated and measured angular diameters agreed, we plotted photometrically-estimated versus interferometrically-measured angular diameters in Figure~\ref{calcobs}, and Figure~\ref{radii} shows a similar plot for physical radii. The angular diameters determined using $K$-band photometry from 2MASS show generally higher errors in Figure~\ref{calcobs} than the diameters determined using TMSS photometry. This plot clearly shows the advantage of measuring angular diameters interferometrically, as the errors are significantly smaller than the photometric estimates in all cases. There is an even scatter around the 1:1 ratio line, and all but two stars are within 1$\sigma$ of the line. 

The outliers in both Figures~\ref{calcobs} and \ref{radii} are HD~118904 and HD~157681. Neither star shows any sign of binarity in the literature, and the SEDs created using the $T_{\rm eff}$ and log~$g$ based on their spectral type and \citet{cox00} do not show any excess in the infrared wavelengths that would suggest a low-mass stellar companion or a circumstellar disk. In both cases, the problem may lie with the calibrator stars chosen. HD~157681 was observed using the calibrator HD~158460, and though the latter has a small estimated diameter (0.268$\pm$0.016 mas) and its SED shows no excess flux in the infrared that would indicate a low-mass stellar companion or circumstellar disk, HD~157681 was the only star observed with that calibrator and there could be an unseen companion that is not taken into account when estimating the star's diameter. Future observations of HD~157681 with different calibrators will make the situation clearer.

HD~118904 was observed using HD~124063 as a calibrator, and the same calibrator was used to observe the target star HD~113049 along with the second calibrator HD~107193. When the data were calibrated separately for HD~113049, the diameters showed a 0.08 mas difference, which is on the order of an 8$\%$ change. If HD~118904's diameter is reduced by 8$\%$, the data point is within errors on the 1:1 ratio line for both plots in Figures~\ref{calcobs} and \ref{radii}. Because this is the case, only HD~107193 was used in the calibration of HD~113049's data, and the angular diameter, radius, and $T_{\rm eff}$ listed in Table~\ref{inf} are based on those data alone.

Figure \ref{radii} shows that while a fair number of photometric and interferometric radii agree very well, there are some that show slight discrepancies, notwithstanding the error bars. This could be due to a few different effects. First, the photometrically-determined radii depend on temperature estimates that may not be correct. If the star is highly active or there is a very faint companion, these could affect the temperature and therefore radii estimates\footnote{If a second star is present and is more than $\sim$2.5 magnitudes fainter than the host star, the effects of the secondart star will be not seen in interferometric observations and would therefore have no effect on the angular diameter or physical radii measurements.}. Second, the limb-darkening law used to determine interferometric diameters and radii may not take certain stellar features into account, such as starspots or extremely active regions. This would not be a large effect because even altering the limb-darkening coefficient $\mu_{\lambda}$ by 20$\%$ changes the limb-darkened angular diameter by an average of 0.7$\%$. Third, the differences may be due to changes in the stars' convections zones, because as the star evolves the convection zone gets deeper. Convection is not well modeled, which may lead to errors in the photometric radii estimates.

\begin{figure}[h]
\includegraphics[width=1.0\textwidth]{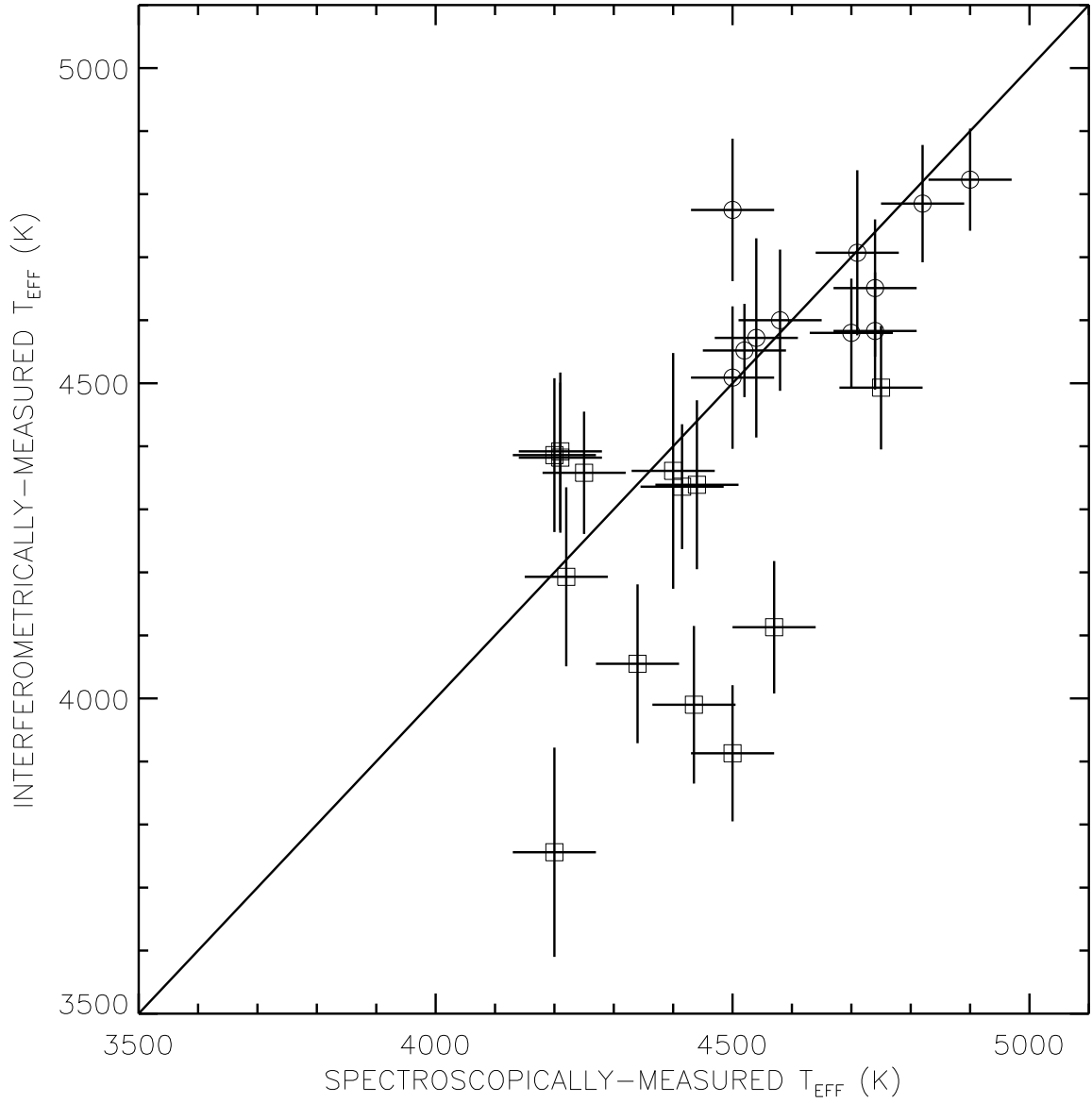}
\caption{A comparison of spectroscopically- and interferometrically-measured effective temperatures. The symbols are the same as listed in Figure~\ref{calcobs}.}
  \label{tcompare}
\end{figure}

We also plotted the interferometrically-measured $T_{\rm eff}$ versus those derived spectroscopically in Figure~\ref{tcompare}. There is some scatter off the 1:1 ratio line, particularly for the cooler stars. The errors in $T_{\rm eff}$ do not show a trend with log~$g$, diameter, radius, $(V-K)$ color, distance, spectral type, metallicity, or bolometric correction. The discrepancies may be due to the inherent properties of the methods used to measure $T_{\rm eff}$. Spectroscopic values are based on Fe\,I and Fe\,II lines and measure the $T_{\rm eff}$ in the part of the atmosphere where those lines are present, while interferometry calculates the overall $T_{\rm eff}$ of the star using the measured diameter. It has been surmised that atmospheric models of K giant stars in the near-ultraviolet band are missing a source of thermal extinction, which would also affect the $T_{\rm eff}$ measurements \citep{2009ApJ...691.1634S}.

Our next step will be to determine the oscillation frequencies of these stars so that we can compare the true masses of these stars with those estimated using evolutionary models. 

\acknowledgments

Many thanks to Douglas Gies for his help and advice. The CHARA Array is funded by the National Science Foundation through NSF grant AST-0606958 and by Georgia State University through the College of Arts and Sciences, and STR acknowledges partial support by NASA grant NNH09AK731. We are also grateful to the user support group of the Alfred-Jensch telescope.
This research has made use of the SIMBAD database, operated at CDS,
Strasbourg, France.  This publication makes use of data products from
the Two Micron All Sky Survey, which is a joint project of the
University of Massachusetts and the Infrared Processing and Analysis
Center/California Institute of Technology, funded by the National
Aeronautics and Space Administration and the National Science
Foundation.


\begin{deluxetable}{llllrcccrrr}
\rotate
\tablewidth{0pc}
\tabletypesize{\scriptsize}
\tablecaption{Observed and Spectroscopic Properties of the K Giants. \label{spec}}

\tablehead{\colhead{Target} & \colhead{$V$} & \colhead{$K$}  & \colhead{Spec}   & \colhead{$\pi$} & \colhead{$T_{\rm eff}$} & \colhead{log~$g$}  & \colhead{[Fe/H]} & \colhead{$\theta_{\rm estimate}$} & \colhead{$R_{\rm estimate}$} & \colhead{$M_{\rm estimate}$} \\ 
           \colhead{HD}       & \colhead{mag} & \colhead{mag}  & \colhead{Type}   & \colhead{(mas)} & \colhead{$\pm$70 K}   & \colhead{$\pm$0.2} & \colhead{$\pm$0.5 dex}    & \colhead{(mas)}             & \colhead{($R_\odot$)}        & \colhead{($M_\odot$)} }
\startdata
32518  & 6.41 & 3.91$\; \pm \;$0.04$^a$ & K1 III & 8.29$\; \pm \;$0.58  & 4580 & 2.0 & -0.15 & 0.84$\; \pm \;$0.05 & 10.9$\; \pm \;$1.0 & 1.1$\; \pm \;$0.2 \\
60294  & 5.92 & 3.55$\; \pm \;$0.22$^a$ & K2 III & 12.24$\; \pm \;$0.39 & 4520 & 2.4 & +0.02 & 0.97$\; \pm \;$0.31 & 8.5$\; \pm \;$2.7 & 1.2$\; \pm \;$0.1 \\
73108  & 4.60 & 1.92$\; \pm \;$0.07$^b$ & K1 III & 12.74$\; \pm \;$0.26 & 4415 & 1.8 & -0.25 & 2.17$\; \pm \;$0.22 & 18.3$\; \pm \;$1.9 & 1.2$\; \pm \;$0.2 \\
102328 & 5.29 & 2.55$\; \pm \;$0.06$^b$ & K3 III & 15.13$\; \pm \;$0.30 & 4250 & 1.9 & +0.09 & 1.64$\; \pm \;$0.14 & 11.6$\; \pm \;$1.0 & 1.1$\; \pm \;$0.1 \\
103605 & 5.84 & 3.10$\; \pm \;$0.30$^a$ & K1 III & 10.54$\; \pm \;$0.37 & 4740 & 2.8 & -0.07 & 1.27$\; \pm \;$0.54 & 12.9$\; \pm \;$5.5 & 1.1$\; \pm \;$0.2 \\
106574 & 5.71 & 2.94$\; \pm \;$0.08$^b$ & K2 III & 7.00$\; \pm \;$0.28  & 4570 & 2.2 & -0.31 & 1.38$\; \pm \;$0.16 & 21.1$\; \pm \;$2.6 & 1.6$\; \pm \;$0.2 \\
113049 & 6.00 & 3.66$\; \pm \;$0.31$^a$ & K0 III & 6.02$\; \pm \;$0.37  & 4740 & 2.2 & -0.18 & 0.92$\; \pm \;$0.41 & 16.4$\; \pm \;$7.3 & 2.2$\; \pm \;$0.3 \\
118904 & 5.51 & 2.69$\; \pm \;$0.07$^b$ & K2 III & 7.93$\; \pm \;$0.24  & 4500 & 2.2 & -0.18 & 1.55$\; \pm \;$0.16 & 21.1$\; \pm \;$2.2 & 1.4$\; \pm \;$0.2 \\
136726 & 5.01 & 1.92$\; \pm \;$0.05$^b$ & K4 III & 8.19$\; \pm \;$0.19  & 4340 & 1.6 & +0.04 & 2.33$\; \pm \;$0.17 & 30.5$\; \pm \;$2.4 & 2.0$\; \pm \;$0.2 \\
137443 & 5.79 & 2.74$\; \pm \;$0.06$^b$ & K4 III & 8.86$\; \pm \;$0.22  & 4435 & 2.6 & -0.03 & 1.58$\; \pm \;$0.14 & 19.2$\; \pm \;$1.7 & 1.4$\; \pm \;$0.2 \\
138265 & 5.88 & 2.38$\; \pm \;$0.04$^b$ & K5 III & 5.11$\; \pm \;$0.31  & 4200 & 2.4 & -0.07 & 2.02$\; \pm \;$0.12 & 42.5$\; \pm \;$3.6 & 1.5$\; \pm \;$0.2 \\
139357 & 5.97 & 3.41$\; \pm \;$0.32$^a$ & K4 III & 8.47$\; \pm \;$0.30  & 4700 & 2.9 & -0.13 & 1.07$\; \pm \;$0.49 & 13.6$\; \pm \;$6.2 & 1.3$\; \pm \;$0.2 \\
150010 & 6.28 & 3.18$\; \pm \;$0.38$^a$ & K2 III & 6.95$\; \pm \;$0.43  & 4540 & 2.8 & -0.02 & 1.31$\; \pm \;$0.71 & 20.2$\; \pm \;$11.1 & 1.4$\; \pm \;$0.3 \\
152812 & 6.00 & 2.83$\; \pm \;$0.09$^b$ & K2 III & 4.97$\; \pm \;$0.45  & 4220 & 1.4 & -0.42 & 1.55$\; \pm \;$0.20 & 33.5$\; \pm \;$5.3 & 1.1$\; \pm \;$0.1 \\
157681 & 5.67 & 2.19$\; \pm \;$0.05$^b$ & K5 III & 5.23$\; \pm \;$0.27  & 4400 & 1.6 & -0.23 & 2.20$\; \pm \;$0.16 & 45.2$\; \pm \;$4.1 & 1.7$\; \pm \;$0.3 \\
160290 & 5.36 & 2.67$\; \pm \;$0.07$^b$ & K1 III & 9.23$\; \pm \;$0.21  & 4750 & 2.7 & -0.17 & 1.54$\; \pm \;$0.16 & 17.9$\; \pm \;$1.9 & 2.0$\; \pm \;$0.3 \\
167042 & 5.98 & 3.55$\; \pm \;$0.24$^a$ & K1 III & 19.91$\; \pm \;$0.26 & 4820 & 2.9 & -0.08 & 0.98$\; \pm \;$0.33 & 5.3$\; \pm \;$1.8 & 1.2$\; \pm \;$0.1 \\
170693 & 4.83 & 1.95$\; \pm \;$0.05$^b$ & K1.5 III & 10.36$\; \pm \;$0.20 & 4200 & 1.0 & -0.46 & 2.21$\; \pm \;$0.16 & 22.9$\; \pm \;$1.7 & 1.0$\; \pm \;$0.1 \\
175823 & 6.22 & 3.57$\; \pm \;$0.32$^a$ & K5 III & 5.63$\; \pm \;$0.28  & 4500 & 2.1 & -0.12 & 1.01$\; \pm \;$0.46 & 19.2$\; \pm \;$8.7 & 1.7$\; \pm \;$0.2 \\
176408 & 5.66 & 3.00$\; \pm \;$0.27$^a$ & K1 III & 11.81$\; \pm \;$0.27 & 4500 & 2.3 & -0.06 & 1.31$\; \pm \;$0.50 & 12.0$\; \pm \;$4.6 & 1.1$\; \pm \;$0.2 \\
186815 & 6.28 & 4.32$\; \pm \;$0.25$^a$ & K2 III & 12.86$\; \pm \;$0.39 & 4900 & 2.5 & -0.32 & 0.63$\; \pm \;$0.23 & 5.3$\; \pm \;$1.9 & 1.2$\; \pm \;$0.1 \\
192781 & 5.79 & 2.33$\; \pm \;$0.07$^b$ & K5 III & 5.62$\; \pm \;$0.23  & 4210 & 2.3 & -0.08 & 2.05$\; \pm \;$0.21 & 39.3$\; \pm \;$4.3 & 1.4$\; \pm \;$0.2 \\
195820 & 6.18 & 3.90$\; \pm \;$0.22$^a$ & K0 III & 8.68$\; \pm \;$0.29  & 4710 & 2.4 & -0.16 & 0.81$\; \pm \;$0.25 & 10.1$\; \pm \;$3.2 & 1.0$\; \pm \;$0.2 \\
200205 & 5.51 & 2.25$\; \pm \;$0.06$^b$ & K4 III & 5.30$\; \pm \;$0.24  & 4210 & 1.6 & -0.28 & 2.06$\; \pm \;$0.18 & 41.7$\; \pm \;$4.1 & 1.3$\; \pm \;$0.2 \\
214868 & 4.48 & 1.41$\; \pm \;$0.07$^b$ & K2 III & 9.80$\; \pm \;$0.26  & 4440 & 2.1 & -0.18 & 2.93$\; \pm \;$0.30 & 32.1$\; \pm \;$3.4 & 1.8$\; \pm \;$0.2 \\
\enddata
\tablecomments{$^a$\emph{2MASS All-Sky Catalog of Point Sources} \citep{2003tmc..book.....C}; $^b$\emph{Two-Micron Sky Survey} \citep{1969tmss.book.....N}; $V$ magnitudes are from Mermilliod (1991) and spectral types are from the \emph{SIMBAD Astronomical Database}; parallaxes ($\pi$) are from \citet{2007hnrr.book.....V}; $T_{\rm eff}$, log~$g$, [Fe/H] from \citet{doellinger08}; $\theta_{\rm estimate}$ and $R_{\rm estimate}$ were determined photometrically, and $M_{\rm estimate}$ is from the PARAM Stellar Model \citep{2006A&A...458..609D}.}
\end{deluxetable}

\clearpage


\begin{deluxetable}{ccrcccccc}
\tablewidth{0pc}
\tablecaption{Observing Log and Calibrator Stars' Basic Parameters.\label{observations}}
\tablehead{\multicolumn{5}{c}{Observing Log} & \multicolumn{1}{c}{ } & \multicolumn{3}{c}{Calibrator Information} \\
\cline{1-5} \cline{7-9} \\
 \colhead{Target} & \colhead{Calibrator} & \colhead{Baseline$^\dagger$} & \colhead{Date} & \colhead{$\#$} & \colhead{ } & \colhead{$T_{\rm eff}$$^\ddagger$} & \colhead{log~$g$$^\ddagger$} & \colhead{$\theta_{\rm LD}$$^\ast$} \\
\colhead{HD} & \colhead{HD}        & \colhead{(max. length)}                  & \colhead{(UT)} & \colhead{Obs} & \colhead{ } & \colhead{(K)} & \colhead{(cm s$^{-2}$)} & \colhead{(mas)} \\ }
\startdata
32518 & 31675  & S1-E1 (331 m) & 2007/11/14 & 9 & & 6310 & 4.39 & 0.401$\pm$0.015 \\
60294 & 63332  & S1-E1 (331 m) & 2009/04/23 & 5 & & 6310 & 4.19 & 0.431$\pm$0.014 \\
      & 69548  &               &            & 5 & & 6761 & 4.31 & 0.402$\pm$0.018 \\
73108 & 69548  & E2-W2 (156 m) & 2008/05/09 & 5 & & 6761 & 4.31 & 0.402$\pm$0.018 \\
102328 & 98673  & S1-E1 (331 m) & 2009/04/24 & 3 & & 8128 & 4.21 & 0.220$\pm$0.010 \\
       & 108954 &               &            & 2 & & 6026 & 4.34 & 0.452$\pm$0.021 \\
103605 & 108954 & S1-E1 (331 m) & 2009/04/22 & 4 & & 6026 & 4.34 & 0.452$\pm$0.021 \\
       & 98673  &               & 2009/04/24 & 3 & & 8128 & 4.21 & 0.220$\pm$0.010 \\
       & 108954 &               &            & 3 & & 6026 & 4.34 & 0.452$\pm$0.021 \\
106574 & 107193 & E2-W2 (156 m) & 2008/06/29 & 6 & & 8710 & 3.93 & 0.315$\pm$0.030 \\
113049 & 107193 & S1-E1 (331 m) & 2009/04/23 & 8 & & 8710 & 3.93 & 0.315$\pm$0.030 \\
       & 124063 &               &            & 5 & & 7740 & 4.29 & 0.232$\pm$0.010 \\
118904 & 124063 & E2-W2 (156 m) & 2008/06/29 & 6 & & 7740 & 4.29 & 0.232$\pm$0.010 \\
136726 & 145454 & E2-W2 (156 m) & 2008/05/09 & 6 & & 9772 & 4.13 & 0.268$\pm$0.015 \\
137443 & 145454 & E2-W2 (156 m) & 2008/05/09 & 6 & & 9772 & 4.13 & 0.268$\pm$0.015 \\
138265 & 145454 & E2-W2 (156 m) & 2008/05/09 & 4 & & 9772 & 4.13 & 0.268$\pm$0.015 \\
       &        &               & 2008/05/11 & 3 & & & & \\
139357 & 132254 & S1-E1 (331 m) & 2007/09/14 & 4 & & 6310 & 4.27 & 0.521$\pm$0.015 \\
       &        &               & 2007/09/15 & 3 & & & & \\
150010 & 145454 & E2-W2 (156 m) & 2008/06/29 & 6 & & 9772 & 4.13 & 0.268$\pm$0.015 \\
       & 149681 & S1-E1 (331 m) & 2008/07/17 & 4 & & 7586 & 4.23 & 0.368$\pm$0.012 \\
152812 & 149303 & S1-E1 (331 m) & 2009/04/20 & 4 & & 8511 & 4.10 & 0.288$\pm$0.011 \\
       & 151044 &               &            & 5 & & 6166 & 4.38 & 0.380$\pm$0.008 \\
157681 & 158460 & S1-E1 (331 m) & 2007/09/14 & 5 & & 9000 & 4.19 & 0.268$\pm$0.016 \\
160290 & 158414 & S1-E1 (331 m) & 2009/04/24 & 6 & & 8000 & 4.24 & 0.295$\pm$0.012 \\
       & 161693 &               &            & 4 & & 9000 & 4.19 & 0.258$\pm$0.015 \\
167042 & 161693 & S1-E1 (331 m) & 2007/09/15 & 8 & & 9000 & 4.19 & 0.258$\pm$0.015 \\
170693 & 172569 & W1-S2 (249 m) & 2007/09/03 & 4 & & 7413 & 3.98 & 0.309$\pm$0.013 \\
175823 & 172728 & S1-E1 (331 m) & 2009/04/23 & 4 & & 9790 & 4.14 & 0.236$\pm$0.020 \\
       & 178207 &               &            & 6 & & 9790 & 4.14 & 0.271$\pm$0.015 \\
176408 & 172728 & S1-E1 (331 m) & 2009/04/23 & 4 & & 9790 & 4.14 & 0.236$\pm$0.020 \\
       & 178207 &               &            & 6 & & 9790 & 4.14 & 0.271$\pm$0.015 \\
186815 & 186760 & S1-E1 (331 m) & 2009/04/24 & 7 & & 6026 & 3.90 & 0.432$\pm$0.019 \\
       & 188793 &               &            & 9 & & 8800 & 4.21 & 0.226$\pm$0.016 \\
192781 & 186760 & S1-E1 (331 m) & 2009/04/24 & 7 & & 6026 & 3.90 & 0.432$\pm$0.019 \\
       & 188793 &               &            & 9 & & 8800 & 4.21 & 0.226$\pm$0.016 \\
195820 & 184960 & S1-E1 (331 m) & 2007/11/14 & 4 & & 6457 & 4.33 & 0.492$\pm$0.019 \\
200205 & 197950 & W1-S2 (249 m) & 2007/09/03 & 8 & & 7762 & 4.30 & 0.349$\pm$0.014 \\
214868 & 211211 & E2-W2 (156 m) & 2008/06/29 & 4 & & 9333 & 4.17 & 0.249$\pm$0.015 \\
       &        & S1-E1 (331 m) & 2008/07/01 & 3 & & & & \\
\enddata
\tablecomments{$^\dagger$The three arms of the Array are denoted by their cardinal directions: ``S'' is south, ``E'' is east, and ``W'' is west. Each arm bears two telescopes, numbered ``1'' for the telescope farthest from the beam combining laboratory and ``2'' for the telescope closer to the lab. \\
$^\ddagger$All $T_{\rm eff}$ and log~$g$ values are from \citet{all99} except for HD 124063, HD 158414, HD 158460, HD 161693, HD 172728, HD 178207, and HD 188793, which are from \citet{cox00} and were based on their spectral types as listed in the \emph{SIMBAD Astronomical Database}. \\ 
$^\ast$In calculating $\theta_{\rm LD}$ as described in $\S$3, the $UBV$ values were from \citet{Mermilliod} except for HD~149303 \citep{1997yCat.1239....0E}, and HD~151044 and HD~184960 \citep{1978A&AS...34..477M}; all $RI$ values were from \citet{2003AJ....125..984M} except for HD~151044 and HD~184960 \citep{1978A&AS...34..477M}; and all $JHK$ values were from \citet{2003tmc..book.....C}.}
\end{deluxetable}



\begin{deluxetable}{ccccccc}
\tablewidth{0pc}
\tablecaption{K Giants' Calibrated Visibilities.\label{calib_visy}}

\tablehead{\colhead{Target} & \colhead{Calib} &  \colhead{ } & \colhead{$B$} & \colhead{$\Theta$} & \colhead{ } & \colhead{ } \\
\colhead{HD} & \colhead{HD} & \colhead{MJD} & \colhead{(m)} & \colhead{(deg)} & \colhead{$V_c$} & \colhead{$\sigma V_c$} \\ }
\startdata
32518 & 31675 & 54418.238 & 230.84 & 200.1 & 0.755 & 0.067 \\
      & & 54418.244 & 233.56 & 201.8 & 0.794 & 0.071 \\
      & &  54418.250 & 236.48 & 203.6 & 0.834 & 0.070 \\
      & &  54418.256 & 239.18 & 205.3 & 0.843 & 0.074 \\
      & &  54418.261 & 241.66 & 206.9 & 0.751 & 0.061 \\
      & &  54418.267 & 244.20 & 208.6 & 0.743 & 0.053 \\
      & &  54418.274 & 246.86 & 210.3 & 0.776 & 0.059 \\
      & &  54418.280 & 249.36 & 212.0 & 0.741 & 0.065 \\
      & &  54418.286 & 251.81 & 213.8 & 0.732 & 0.053 \\
60294 & 63332 & 54944.176 & 319.48 & 94.9 & 0.444 & 0.045 \\
      & &  54944.184 & 319.22 & 96.9 & 0.451 & 0.054 \\
      & &  54944.192 & 318.88 & 98.9 & 0.474 & 0.036 \\
      & &  54944.201 & 318.44 & 100.8 & 0.481 & 0.035 \\
      & &  54944.208 & 317.93 & 102.7 & 0.449 & 0.058 \\
      & 69548 & 54944.160 & 319.72 & 91.0 & 0.526 & 0.057 \\
      & &  54944.168 & 319.64 & 92.9 & 0.448 & 0.057 \\
      & &  54944.192 & 318.88 & 98.9 & 0.485 & 0.040 \\
      & &  54944.201 & 318.44 & 100.8 & 0.505 & 0.052 \\
      & &  54944.208 & 317.93 & 102.7 & 0.455 & 0.053 \\
73108 & 69548 & 54595.216 & 155.95 & 254.7 & 0.411 & 0.051 \\
      & &  54595.226 & 155.88 & 258.0 & 0.446 & 0.034 \\
      & &  54595.235 & 155.83 & 261.1 & 0.436 & 0.043 \\
      & &  54595.244 & 155.80 & 264.1 & 0.460 & 0.057 \\
      & &  54595.257 & 155.77 & 268.4 & 0.430 & 0.092 \\
102328 & 98673 & 54945.239 & 314.63 & 248.9 & 0.086 & 0.011 \\
      & &  54945.252 & 316.18 & 252.0 & 0.088 & 0.009 \\
      & &  54945.290 & 319.04 & 261.0 & 0.073 & 0.011 \\
      & 108954 & 54945.239 & 314.63 & 248.9 & 0.100 & 0.012 \\
      & &  54945.264 & 317.31 & 254.8 & 0.095 & 0.012 \\
103605 & 108954 & 54943.375 & 317.53 & 99.1 & 0.437 & 0.027 \\
      & &  54943.382 & 317.18 & 100.6 & 0.442 & 0.032 \\
      & &  54943.388 & 316.77 & 102.2 & 0.457 & 0.038 \\
      & &  54943.394 & 316.37 & 103.5 & 0.409 & 0.029 \\
      & 98673 & 54945.242 & 312.54 & 248.0 & 0.445 & 0.048 \\
      & &  54945.255 & 314.17 & 251.0 & 0.410 & 0.040 \\
      & &  54945.280 & 316.51 & 256.9 & 0.446 & 0.062 \\
      & 108954 & 54945.267 & 315.40 & 253.8 & 0.465 & 0.069 \\
      & &  54945.280 & 316.51 & 256.9 & 0.489 & 0.054 \\
      & &  54945.293 & 317.37 & 260.2 & 0.449 & 0.044 \\
106574 & 107193 & 54646.187 & 155.91 & 241.7 & 0.699 & 0.099 \\
       & &  54646.196 & 155.99 & 244.8 & 0.698 & 0.131 \\
       & &  54646.205 & 156.06 & 247.8 & 0.741 & 0.098 \\
       & &  54646.214 & 156.11 & 250.8 & 0.680 & 0.086 \\
       & &  54646.223 & 156.14 & 253.9 & 0.732 & 0.085 \\
       & &  54646.234 & 156.18 & 257.5 & 0.701 & 0.081 \\
113049 & 107193 & 54944.362 & 272.32 & 265.1 & 0.655 & 0.059 \\
       & &  54944.370 & 272.47 & 267.4 & 0.630 & 0.051 \\
       & &  54944.378 & 272.53 & 269.5 & 0.692 & 0.070 \\
       & &  54944.386 & 272.50 & 91.9 & 0.670 & 0.052 \\
       & &  54944.394 & 272.37 & 94.4 & 0.587 & 0.049 \\
       & &  54944.403 & 272.12 & 96.8 & 0.605 & 0.049 \\
       & &  54944.411 & 271.80 & 99.0 & 0.633 & 0.076 \\
       & &  54944.419 & 271.38 & 101.3 & 0.696 & 0.071 \\
       & 124063 & 54944.362 & 272.32 & 265.1 & 0.611 & 0.059 \\
       & &  54944.370 & 272.47 & 267.4 & 0.542 & 0.035 \\
       & &  54944.378 & 272.53 & 269.5 & 0.602 & 0.048 \\
       & &  54944.411 & 271.80 & 99.0 & 0.581 & 0.070 \\
       & &  54944.419 & 271.38 & 101.3 & 0.656 & 0.058 \\
118904 & 124063 & 54646.251 & 155.81 & 244.4 & 0.574 & 0.074 \\
       & &  54646.260 & 155.89 & 247.3 & 0.567 & 0.069 \\
       & &  54646.268 & 155.95 & 250.1 & 0.589 & 0.060 \\
       & &  54646.278 & 156.01 & 253.3 & 0.512 & 0.064 \\
       & &  54646.288 & 156.05 & 256.6 & 0.583 & 0.070 \\
       & &  54646.297 & 156.08 & 259.6 & 0.562 & 0.088 \\
136726 & 145454 & 54595.294 & 147.57 & 189.4 & 0.442 & 0.055 \\
       & &  54595.307 & 148.79 & 193.7 & 0.425 & 0.045 \\
       & &  54595.315 & 149.53 & 196.5 & 0.468 & 0.054 \\
       & &  54595.325 & 150.30 & 199.6 & 0.421 & 0.056 \\
       & &  54595.336 & 151.17 & 203.4 & 0.436 & 0.062 \\
       & &  54595.346 & 151.80 & 206.5 & 0.409 & 0.053 \\
137443 & 145454 & 54595.385 & 155.65 & 219.1 & 0.673 & 0.082 \\
       & &  54595.394 & 155.90 & 222.1 & 0.631 & 0.083 \\
       & &  54595.404 & 156.07 & 225.1 & 0.616 & 0.063 \\
       & &  54595.415 & 156.20 & 228.5 & 0.605 & 0.068 \\
       & &  54595.430 & 156.26 & 233.6 & 0.656 & 0.077 \\
       & &  54595.440 & 156.24 & 236.6 & 0.664 & 0.048 \\
138265 & 145454 & 54595.455 & 155.83 & 240.4 & 0.542 & 0.081 \\
       & &  54595.466 & 155.64 & 243.8 & 0.578 & 0.130 \\
       & &  54595.476 & 155.45 & 247.1 & 0.494 & 0.085 \\
       & &  54595.488 & 155.23 & 251.0 & 0.467 & 0.070 \\
       & &  54597.467 & 155.50 & 246.2 & 0.559 & 0.079 \\
       & &  54597.477 & 155.33 & 249.3 & 0.500 & 0.061 \\
       & &  54597.486 & 155.16 & 252.4 & 0.478 & 0.061 \\
139357 & 132254 & 54357.149 & 320.57 & 102.8 & 0.450 & 0.070 \\
       & &  54357.155 & 320.14 & 104.2 & 0.460 & 0.045 \\
       & &  54357.161 & 319.66 & 105.6 & 0.487 & 0.063 \\
       & &  54357.167 & 319.12 & 107.1 & 0.491 & 0.066 \\
       & &  54358.151 & 320.24 & 103.9 & 0.460 & 0.030 \\
       & &  54358.157 & 319.77 & 105.3 & 0.415 & 0.034 \\
       & &  54358.162 & 319.27 & 106.7 & 0.429 & 0.049 \\
150010 & 145454 & 54646.318 & 154.39 & 226.3 & 0.785 & 0.122 \\
       & &  54646.327 & 154.64 & 229.1 & 0.836 & 0.103 \\
       & &  54646.335 & 154.85 & 231.9 & 0.823 & 0.085 \\
       & &  54646.345 & 155.07 & 235.1 & 0.835 & 0.113 \\
       & &  54646.354 & 155.24 & 238.2 & 0.862 & 0.125 \\
       & &  54646.363 & 155.39 & 241.1 & 0.822 & 0.083 \\
       & 149681 &  54664.392 & 273.81 & 117.3 & 0.583 & 0.086 \\
       & &  54664.403 & 272.04 & 120.2 & 0.589 & 0.096 \\
       & &  54664.413 & 270.21 & 123.0 & 0.640 & 0.080 \\
       & &  54664.423 & 268.16 & 125.8 & 0.724 & 0.125 \\
152812 & 149303 & 54941.490 & 327.27 & 256.0 & 0.152 & 0.014 \\
       & &  54941.499 & 327.63 & 257.9 & 0.164 & 0.015 \\
       & &  54941.507 & 327.94 & 259.9 & 0.154 & 0.010 \\
       & &  54941.516 & 328.18 & 261.9 & 0.148 & 0.015 \\
       & 151044 & 54941.481 & 326.75 & 253.8 & 0.158 & 0.017 \\
       & &  54941.490 & 327.27 & 256.0 & 0.144 & 0.014 \\
       & &  54941.499 & 327.63 & 257.9 & 0.168 & 0.015 \\
       & &  54941.507 & 327.94 & 259.9 & 0.157 & 0.011 \\
       & &  54941.516 & 328.18 & 261.9 & 0.154 & 0.016 \\
157681 & 158460 & 54357.211 & 321.92 & 99.9 & 0.056 & 0.004 \\
       & &  54357.221 & 321.36 & 102.3 & 0.056 & 0.005 \\
       & &  54357.231 & 320.64 & 104.8 & 0.057 & 0.005 \\
       & &  54357.241 & 319.83 & 107.1 & 0.060 & 0.004 \\
       & &  54357.251 & 318.83 & 109.4 & 0.060 & 0.010 \\
160290 & 158414 & 54945.330 & 279.04 & 219.2 & 0.267 & 0.020 \\
       & &  54945.339 & 283.87 & 220.7 & 0.280 & 0.030 \\
       & &  54945.348 & 288.23 & 222.2 & 0.247 & 0.030 \\
       & &  54945.356 & 292.52 & 223.8 & 0.250 & 0.037 \\
       & &  54945.366 & 296.96 & 225.7 & 0.217 & 0.042 \\
       & &  54945.375 & 300.60 & 227.4 & 0.172 & 0.022 \\
       & 161693 & 54945.348 & 288.23 & 222.2 & 0.225 & 0.026 \\
       & &  54945.356 & 292.52 & 223.8 & 0.228 & 0.027 \\
       & &  54945.366 & 296.96 & 225.7 & 0.183 & 0.021 \\
       & &  54945.375 & 300.60 & 227.4 & 0.167 & 0.017 \\
167042 & 161693 & 54358.232 & 321.20 & 97.5 & 0.584 & 0.037 \\
       & &  54358.238 & 320.96 & 99.0 & 0.551 & 0.036 \\
       & &  54358.243 & 320.68 & 100.3 & 0.507 & 0.036 \\
       & &  54358.249 & 320.34 & 101.7 & 0.524 & 0.030 \\
       & &  54358.255 & 319.96 & 103.1 & 0.571 & 0.036 \\
       & &  54358.261 & 319.53 & 104.5 & 0.612 & 0.037 \\
       & &  54358.267 & 319.05 & 105.9 & 0.591 & 0.041 \\
       & &  54358.273 & 318.48 & 107.4 & 0.627 & 0.050 \\
170693 & 172569 & 54346.303 & 187.40 & 183.8 & 0.373 & 0.042 \\
       & &  54346.311 & 183.87 & 186.6 & 0.343 & 0.049 \\
       & &  54346.321 & 179.32 & 190.2 & 0.358 & 0.037 \\
       & &  54346.332 & 174.70 & 193.9 & 0.457 & 0.042 \\
175823 & 172728 & 54944.471 & 297.24 & 232.9 & 0.499 & 0.044 \\
       & &  54944.482 & 300.30 & 235.4 & 0.480 & 0.064 \\
       & &  54944.493 & 303.23 & 238.1 & 0.553 & 0.065 \\
       & &  54944.505 & 305.73 & 240.8 & 0.533 & 0.056 \\
       & 178207 & 54944.442 & 287.69 & 226.5 & 0.633 & 0.071 \\
       & &  54944.454 & 291.73 & 229.0 & 0.667 & 0.053 \\
       & &  54944.471 & 297.24 & 232.9 & 0.590 & 0.052 \\
       & &  54944.482 & 300.30 & 235.4 & 0.576 & 0.082 \\
       & &  54944.493 & 303.23 & 238.1 & 0.569 & 0.060 \\
       & &  54944.505 & 305.73 & 240.8 & 0.580 & 0.045 \\
176408 & 172728 & 54944.473 & 296.67 & 232.9 & 0.409 & 0.043 \\
       & &  54944.484 & 299.83 & 235.5 & 0.409 & 0.047 \\
       & &  54944.496 & 302.66 & 238.1 & 0.436 & 0.059 \\
       & &  54944.507 & 305.17 & 240.8 & 0.416 & 0.046 \\
       & 178207 & 54944.445 & 287.23 & 226.5 & 0.585 & 0.060 \\
       & &  54944.456 & 291.14 & 228.9 & 0.587 & 0.055 \\
       & &  54944.473 & 296.67 & 232.9 & 0.501 & 0.053 \\
       & &  54944.484 & 299.83 & 235.5 & 0.462 & 0.058 \\
       & &  54944.496 & 302.66 & 238.1 & 0.452 & 0.057 \\
       & &  54944.507 & 305.17 & 240.8 & 0.465 & 0.039 \\
186815 & 186760 & 54945.396 & 248.69 & 209.6 & 0.792 & 0.082 \\
       & &  54945.408 & 256.16 & 212.2 & 0.891 & 0.083 \\
       & &  54945.419 & 262.58 & 214.5 & 0.732 & 0.056 \\
       & &  54945.429 & 268.32 & 216.8 & 0.777 & 0.069 \\
       & &  54945.440 & 273.75 & 219.1 & 0.764 & 0.072 \\
       & &  54945.484 & 292.08 & 228.8 & 0.815 & 0.053 \\
       & &  54945.495 & 295.81 & 231.3 & 0.740 & 0.058 \\
       & 188793 & 54945.396 & 248.69 & 209.6 & 0.749 & 0.098 \\
       & &  54945.408 & 256.16 & 212.2 & 0.929 & 0.111 \\
       & &  54945.419 & 262.58 & 214.5 & 0.760 & 0.082 \\
       & &  54945.429 & 268.32 & 216.8 & 0.699 & 0.075 \\
       & &  54945.440 & 273.75 & 219.1 & 0.742 & 0.069 \\
       & &  54945.461 & 283.52 & 223.8 & 0.783 & 0.058 \\
       & &  54945.473 & 287.96 & 226.3 & 0.778 & 0.040 \\
       & &  54945.484 & 292.08 & 228.8 & 0.761 & 0.044 \\
       & &  54945.495 & 295.81 & 231.3 & 0.726 & 0.057 \\
192781 & 186760 & 54945.400 & 231.04 & 202.6 & 0.225 & 0.027 \\
       & &  54945.411 & 238.31 & 205.2 & 0.202 & 0.017 \\
       & &  54945.422 & 245.12 & 207.9 & 0.174 & 0.012 \\
       & &  54945.432 & 251.29 & 210.4 & 0.172 & 0.014 \\
       & &  54945.443 & 257.29 & 212.9 & 0.140 & 0.012 \\
       & &  54945.487 & 277.78 & 223.2 & 0.078 & 0.005 \\
       & &  54945.498 & 282.14 & 225.9 & 0.062 & 0.004 \\
       & 188793 &  54945.400 & 231.04 & 202.6 & 0.220 & 0.032 \\
       & &  54945.411 & 238.31 & 205.2 & 0.214 & 0.024 \\
       & &  54945.422 & 245.12 & 207.9 & 0.173 & 0.018 \\
       & &  54945.432 & 251.29 & 210.4 & 0.154 & 0.016 \\
       & &  54945.443 & 257.29 & 212.9 & 0.143 & 0.012 \\
       & &  54945.464 & 268.07 & 218.0 & 0.109 & 0.010 \\
       & &  54945.476 & 273.15 & 220.6 & 0.091 & 0.006 \\
       & &  54945.487 & 277.78 & 223.2 & 0.074 & 0.004 \\
       & &  54945.498 & 282.14 & 225.9 & 0.062 & 0.004 \\
195820 & 184960 & 54418.169 & 323.56 & 98.4 & 0.626 & 0.070 \\
       & &  54418.184 & 322.84 & 102.0 & 0.703 & 0.076 \\
       & &  54418.194 & 322.25 & 104.2 & 0.546 & 0.051 \\
       & &  54418.203 & 321.50 & 106.5 & 0.610 & 0.059 \\
200205 & 197950 & 54346.350 & 214.19 & 161.4 & 0.327 & 0.032 \\
       & &  54346.358 & 211.24 & 163.5 & 0.309 & 0.039 \\
       & &  54346.365 & 208.36 & 165.4 & 0.267 & 0.030 \\
       & &  54346.372 & 205.28 & 167.4 & 0.302 & 0.039 \\
       & &  54346.378 & 202.26 & 169.3 & 0.338 & 0.033 \\
       & &  54346.385 & 199.22 & 171.2 & 0.242 & 0.021 \\
       & &  54346.392 & 195.69 & 173.3 & 0.267 & 0.035 \\
       & &  54346.406 & 188.87 & 177.4 & 0.318 & 0.031 \\
214868 & 211211 & 54646.402 & 138.11 & 183.1 & 0.361 & 0.072 \\
       & &  54646.413 & 141.40 & 185.6 & 0.352 & 0.042 \\
       & &  54646.423 & 144.26 & 188.0 & 0.326 & 0.054 \\
       & &  54646.433 & 146.89 & 190.5 & 0.304 & 0.058 \\
       & &  54648.457 & 322.39 & 239.2 & 0.073 & 0.012 \\
       & &  54648.469 & 324.32 & 241.8 & 0.064 & 0.005 \\
       & &  54648.479 & 325.51 & 243.8 & 0.079 & 0.007 \\
\enddata
\tablecomments{The projected baseline position angle ($\Theta$) is calculated to be east of north.}
\end{deluxetable}

\clearpage


\begin{deluxetable}{lcccrcccclc}
\rotate
\tablewidth{0pc}
\tabletypesize{\scriptsize}
\tablecaption{Interferometric Diameter and Effective Temperature Measurements of the K Giants. \label{inf}}

\tablehead{\colhead{Target} &  \colhead{$\theta_{\rm UD,interferometric}$} & \colhead{$\theta_{\rm LD,interferometric}$} & \colhead{$\sigma_{\rm LD}$} & \colhead{$R_{\rm linear}$} & \colhead{ }           & \colhead{ }  & \colhead{$L_{\star}$}         & \colhead{$F_{\rm BOL}$}                      & \colhead{$T_{\rm eff}$} & \colhead{$\sigma_{\rm Teff}$} \\ 
           \colhead{HD}       & \colhead{(mas)}                 & \colhead{(mas)}                 & \colhead{($\%$)} & \colhead{($R_\odot$)}  & \colhead{$A_{\rm V}$} & \colhead{BC} & \colhead{($L_{\odot}$)} & \colhead{(10$^{-8}$ erg s$^{-1}$ cm$^{-2}$)} & \colhead{(K)}           & \colhead{$\%$} }
\startdata
32518  & 0.828$\; \pm \;$0.022 & 0.851$\; \pm \;$0.022 & 3   & 11.04$\; \pm \;$0.77 & 0.06 & 0.43$\; \pm \;$0.08 &  49.2$\; \pm \;$3.6 &  10.8$\; \pm \;$0.9 & 4600$\; \pm \;$112 & 2  \\
60294  & 1.014$\; \pm \;$0.010 & 1.044$\; \pm \;$0.010 & 1   &  9.17$\; \pm \;$0.29 & 0.05 & 0.35$\; \pm \;$0.06 &  32.5$\; \pm \;$1.9  & 15.6$\; \pm \;$1.0 & 4552$\; \pm \;$74 & 2  \\
73108  & 2.161$\; \pm \;$0.019 & 2.225$\; \pm \;$0.020 & 1   & 18.79$\; \pm \;$0.38 & 0.00 & 0.51$\; \pm \;$0.09 & 112.4$\; \pm \;$10.0  & 58.3$\; \pm \;$5.2 & 4336$\; \pm \;$99 & 2 \\
102328 & 1.546$\; \pm \;$0.006 & 1.606$\; \pm \;$0.006 & 0.4 & 11.42$\; \pm \;$0.23 & 0.00 & 0.51$\; \pm \;$0.09 &  42.4$\; \pm \;$3.8  & 31.0$\; \pm \;$2.8 & 4358$\; \pm \;$97 & 2  \\
103605 & 1.066$\; \pm \;$0.009 & 1.098$\; \pm \;$0.010 & 1   & 11.20$\; \pm \;$0.41 & 0.00 & 0.52$\; \pm \;$0.09 &  52.9$\; \pm \;$4.8  & 18.8$\; \pm \;$1.7 & 4651$\; \pm \;$109 & 2  \\
106574 & 1.458$\; \pm \;$0.027 & 1.498$\; \pm \;$0.028 & 2   & 23.02$\; \pm \;$0.92 & 0.00 & 0.54$\; \pm \;$0.10 & 136.6$\; \pm \;$12.7  & 21.4$\; \pm \;$2.0 & 4113$\; \pm \;$105 & 3 \\
113049$^\dagger$ & 0.945$\; \pm \;$0.021 & 0.971$\; \pm \;$0.022 & 2   & 17.35$\; \pm \;$1.07 & 0.00 & 0.35$\; \pm \;$0.06 & 119.7$\; \pm \;$7.2  & 13.9$\; \pm \;$0.9 & 4583$\; \pm \;$93 & 2  \\
118904 & 1.842$\; \pm \;$0.031 & 1.871$\; \pm \;$0.032 & 2   & 25.38$\; \pm \;$0.88 & 0.00 & 0.60$\; \pm \;$0.11 & 136.0$\; \pm \;$14.1 & 27.3$\; \pm \;$2.9 & 3913$\; \pm \;$108 & 3  \\
136726 & 2.264$\; \pm \;$0.020 & 2.336$\; \pm \;$0.020 & 1   & 30.68$\; \pm \;$0.76 & 0.04 & 0.70$\; \pm \;$0.13 & 229.2$\; \pm \;$28.2 & 49.2$\; \pm \;$6.1 & 4055$\; \pm \;$126 & 3  \\
137443 & 1.638$\; \pm \;$0.030 & 1.690$\; \pm \;$0.031 & 2   & 20.51$\; \pm \;$0.62 & 0.06 & 0.68$\; \pm \;$0.12 & 96.1$\; \pm \;$11.5 & 24.1$\; \pm \;$2.9 & 3990$\; \pm \;$125 & 3   \\
138265 & 1.998$\; \pm \;$0.037 & 2.062$\; \pm \;$0.038 & 2   & 43.40$\; \pm \;$2.75 & 0.06 & 0.95$\; \pm \;$0.17 & 337.8$\; \pm \;$57.5 & 28.2$\; \pm \;$4.9 & 3758$\; \pm \;$166 & 4  \\
139357 & 1.040$\; \pm \;$0.012 & 1.073$\; \pm \;$0.013 & 1   & 13.63$\; \pm \;$0.51 & 0.13 & 0.40$\; \pm \;$0.07 &  73.6$\; \pm \;$5.1  & 16.9$\; \pm \;$1.2 & 4580$\; \pm \;$86 & 2   \\
150010 & 0.995$\; \pm \;$0.028 & 1.024$\; \pm \;$0.029 & 3   & 15.84$\; \pm \;$1.08 & 0.04 & 0.70$\; \pm \;$0.13 &  98.9$\; \pm \;$12.2 & 15.3$\; \pm \;$1.9 & 4572$\; \pm \;$158 & 3 \\
152812 & 1.393$\; \pm \;$0.003 & 1.440$\; \pm \;$0.004 & 0.3 & 31.16$\; \pm \;$2.82 & 0.10 & 0.72$\; \pm \;$0.13 & 270.5$\; \pm \;$34.4 & 21.4$\; \pm \;$2.9 & 4193$\; \pm \;$142 & 3  \\
157681 & 1.600$\; \pm \;$0.009 & 1.664$\; \pm \;$0.010 & 1   & 34.22$\; \pm \;$1.78 & 0.04 & 0.94$\; \pm \;$0.17 & 381.7$\; \pm \;$64.4 & 33.4$\; \pm \;$5.7 & 4361$\; \pm \;$187 & 4 \\
160290 & 1.467$\; \pm \;$0.010 & 1.515$\; \pm \;$0.010 & 1   & 17.65$\; \pm \;$0.42 & 0.10 & 0.50$\; \pm \;$0.09 & 114.4$\; \pm \;$9.5  & 31.2$\; \pm \;$2.7 & 4493$\; \pm \;$98 & 2  \\
167042 & 0.898$\; \pm \;$0.017 & 0.922$\; \pm \;$0.018 & 2   &  4.98$\; \pm \;$0.07 & 0.01 & 0.39$\; \pm \;$0.07 &  11.7$\; \pm \;$0.8  & 14.8$\; \pm \;$1.0 & 4785$\; \pm \;$93 & 2  \\
170693 & 1.981$\; \pm \;$0.041 & 2.041$\; \pm \;$0.043 & 2   & 21.19$\; \pm \;$0.60 & 0.02 & 0.59$\; \pm \;$0.11 & 149.7$\; \pm \;$15.3 & 51.4$\; \pm \;$5.3 & 4386$\; \pm \;$122 & 3 \\
175823 & 0.958$\; \pm \;$0.022 & 0.988$\; \pm \;$0.023 & 2   & 18.88$\; \pm \;$1.04 & 0.05 & 0.49$\; \pm \;$0.09 & 132.7$\; \pm \;$11.3  & 13.4$\; \pm \;$1.2 & 4509$\; \pm \;$113 & 3  \\
176408 & 1.092$\; \pm \;$0.022 & 1.125$\; \pm \;$0.023 & 2   & 10.24$\; \pm \;$0.23 & 0.02 & 0.49$\; \pm \;$0.09 &  49.2$\; \pm \;$4.2  & 21.9$\; \pm \;$1.9 & 4775$\; \pm \;$113 & 2  \\
186815 & 0.713$\; \pm \;$0.020 & 0.731$\; \pm \;$0.020 & 3   &  6.11$\; \pm \;$0.25 & 0.02 & 0.21$\; \pm \;$0.04 &  18.2$\; \pm \;$0.7  &  9.6$\; \pm \;$0.4 & 4823$\; \pm \;$81 & 2  \\
192781 & 1.787$\; \pm \;$0.002 & 1.859$\; \pm \;$0.003 & 0.2 & 35.57$\; \pm \;$1.46 & 0.40 & 0.62$\; \pm \;$0.11 & 405.2$\; \pm \;$43.8 & 40.9$\; \pm \;$4.5 & 4342$\; \pm \;$119 & 3  \\
195820 & 0.840$\; \pm \;$0.040 & 0.863$\; \pm \;$0.041 & 5   & 10.69$\; \pm \;$0.62 & 0.07 & 0.33$\; \pm \;$0.06 &  50.6$\; \pm \;$2.8  & 12.2$\; \pm \;$0.7 & 4707$\; \pm \;$131 & 3 \\
200205 & 1.963$\; \pm \;$0.043 & 2.032$\; \pm \;$0.045 & 2   & 41.23$\; \pm \;$2.08 & 0.69 & 0.59$\; \pm \;$0.11 & 569.9$\; \pm \;$58.6 & 51.2$\; \pm \;$5.4 & 4392$\; \pm \;$125 & 3 \\
214868 & 2.721$\; \pm \;$0.020 & 2.731$\; \pm \;$0.024 & 1   & 29.98$\; \pm \;$0.84 & 0.15 & 0.69$\; \pm \;$0.12 & 286.9$\; \pm \;$34.9 & 88.1$\; \pm \;$10.8 & 4339$\; \pm \;$134 & 3 \\
\enddata
\tablecomments{$^\dagger$The angular diameter and subsequent calculations are based on data calibrated using HD~107193 only. See $\S$5 for more details.}
\end{deluxetable}

\clearpage


\clearpage

\clearpage

\clearpage

\clearpage

\clearpage



\end{document}